\documentclass[runningheads]{llncs}
\usepackage{graphicx}
\usepackage{hyperref}

\usepackage{subcaption}
\usepackage{amsmath}
\usepackage{amssymb}
\usepackage{bm}
\usepackage{algorithm}
\usepackage[noend]{algpseudocode}
\usepackage{multirow}
\usepackage{booktabs} 
\usepackage{tabularx}
\usepackage{xcolor}

\definecolor{red}{RGB}{246, 193, 192}
\definecolor{pink}{RGB}{183, 178, 254}
\definecolor{cyan}{RGB}{205, 207, 196}

\begin{document}
\title{Local Citation Recommendation with Hierarchical-Attention Text Encoder and SciBERT-based Reranking}
\titlerunning{Local Citation Recommendation with HAtten and SciBERT }
%
\author{
Nianlong Gu \and
Yingqiang Gao \and
Richard H.R. Hahnloser
}
\authorrunning{Nianlong Gu et al.}
%
\institute{
Institute of Neuroinformatics,
University of Zurich and
ETH Zurich\\
\email{\{nianlong,yingqiang.gao,rich\}@ini.ethz.ch}
} 
\maketitle              
\begin{abstract}

The goal of local citation recommendation is to recommend a missing reference from the local citation context and optionally also from the global context. To balance the tradeoff between speed and accuracy of citation recommendation in the context of a large-scale paper database, a viable approach is to first prefetch a limited number of relevant documents using efficient ranking methods and then to perform a fine-grained reranking using more sophisticated models. In that vein, BM25 has been found to be a tough-to-beat approach to prefetching, which is why recent work has focused mainly on the reranking step. Even so, we explore prefetching with nearest neighbor search among text embeddings constructed by a hierarchical attention network. When coupled with a SciBERT reranker fine-tuned on local citation recommendation tasks, our hierarchical Attention encoder (HAtten) achieves high prefetch recall for a given number of candidates to be reranked. Consequently, our reranker requires fewer prefetch candidates to rerank, yet still achieves state-of-the-art performance on various local citation recommendation datasets such as ACL-200, FullTextPeerRead, RefSeer, and arXiv.

\end{abstract}

\section{Introduction}

Literature discovery, such as finding relevant scientific articles, remains challenging in today’s age of information overflow, largely arising from the exponential growth in both the publication record \cite{hunter2006biomedical} and the underlying vocabulary \cite{herdan1960type}. Assistance to literature discovery can be provided with automatic citation recommendation, whereby a query text without citation serves as the input to a recommendation system and a paper worth citing as its output~\cite{F_rber_2020}. 

\begin{figure}[ht]
    \centering

    \includegraphics[width=\linewidth]{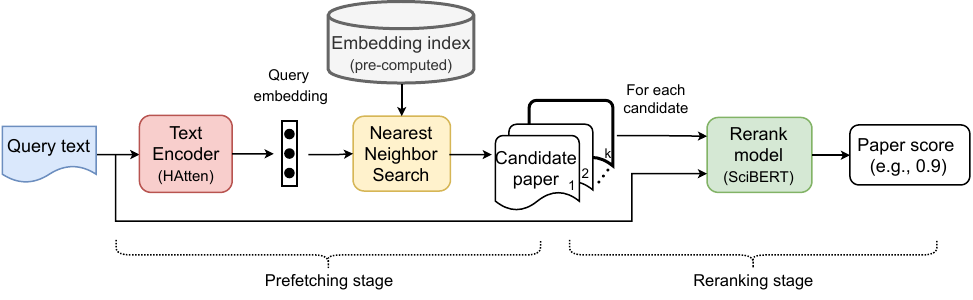}
      
    \caption{Overview of our two-stage local citation recommendation pipeline.}
    \label{fig:pipeline}
\end{figure}

Citation recommendation can be dealt with either as a global retrieval problem \cite{strohman2007recommending,nallapati2008joint,bhagavatula-etal-2018-content} or as a local one \cite{he2010context,huang2012recommending,jeong2019context}. In global citation recommendation, the query text is composed of the title and the abstract of a source paper \cite{bhagavatula-etal-2018-content}. In contrast, in local citation recommendation, the query consist of two sources of contexts \cite{he2010context,medic-snajder-2020-improved}: 1) the text surrounding the citation placeholder with the information of the cited paper removed (the \textbf{local context}); and 2) the title and abstract of the citing paper as the \textbf{global context}. The aim of local citation recommendation is to find the missing paper cited at the placeholder of the local context. In this paper we focus on local citation recommendation.

It is important for a local citation recommendation system to maintain a balance between accuracy (e.g., recall of the target paper among the top K recommended papers) and speed in order to operate efficiently on a large database containing millions of scientific papers. The speed-accuracy tradeoff can be flexibly dealt with using a two-step prefetching-reranking strategy: 1) A fast prefetching model first retrieves a set of candidate papers from the database; 2) a more sophisticated model then performs a fine-grained analysis of scoring candidate papers and reordering them to result in a ranked list of recommendations.
In many recent studies \cite{medic-snajder-2020-improved,10.1109/TASLP.2019.2949925,10.1145/3077136.3080730,10.1145/2600428.2609585}, either (TF-IDF) \cite{ramos2003using} or BM25 \cite{robertson2009probabilistic} were used as the prefetching algorithm, which were neither fine-tuned nor taken into consideration when evaluating the recommendation performance.

In this paper, we propose a novel two-stage local citation recommendation system (Figure \ref{fig:pipeline}). In the prefetching stage, we make use of an embedding-based paper retrieval system, in which a siamese text encoder first pre-computes a vector-based embedding for each paper in the database. The query text is then mapped into the same embedding space to retrieve the $K$ nearest neighbors of the query vector. To encode queries and papers of various lengths in a memory-efficient way, we design a two-layer Hierarchical Attention-based text encoder (HAtten) that first computes paragraph embeddings and then computes from the paragraph embeddings the query and document embeddings using a self-attention mechanism \cite{vaswani2017attention}. In the reranking step, we fine-tune the SciBERT \cite{beltagy2019scibert} to rerank the candidates retrieved by the HAtten prefetching model.

In addition, to cope with the scarceness of large-scale training datasets in many domains, we construct a novel dataset that we distilled from 1.7 million arXiv papers. The dataset consist of 3.2 million local citation sentences along with the title and the abstract of both the citing and the cited papers. Extensive experiments on the arXiv dataset as well as on previous datasets including ACL-200 \cite{medic-snajder-2020-improved},  RefSeer \cite{medic-snajder-2020-improved,10.1145/3077136.3080730}, and FullTextPeerRead \cite{jeong2019context} show that our local citation recommendation system performs better on both prefetching and reranking than the baseline and requires fewer prefetched candidates in the reranking step thanks to higher recall of our prefetching system, which indicates that our system strikes a better speed-accuracy balance.

In total, our main contributions are summarized as follows:
1) We propose a competitive retrieval system consisting of a hierarchical-attention text encoder and a fine-tuned SciBERT reranker.
2) In evaluations of the whole pipeline, we demonstrate a well-balanced tradeoff between speed and accuracy.
3) We release our code and a large-scale scientific paper dataset\footnote{Our code and data are available at \url{https://github.com/nianlonggu/Local-Citation-Recommendation}.} for training and evaluation of production-level local citation recommendation systems.

\section{Related Work}

Local citation recommendation was previously addressed in He et al. \cite{he2010context} in which a non-parametric probabilistic model was proposed to model the relevance between the query and each candidate citation. In recent years, embedding-based approaches \cite{10.1145/3197026.3197059,gokcce2020embedding} have been proposed to more flexibly capture the resemblance between the query and the target according to the cosine distance or the Euclidean distance between their embeddings. Jeong et al. \cite{jeong2019context} proposed a BERT-GCN model in which they used Graph Convolutional Networks \cite{DBLP:conf/iclr/KipfW17} (GCN) and BERT \cite{devlin-etal-2019-bert} to compute for each paper embeddings of the citation graph and the query context, which they fed into a feed-forward network to estimate relevance. The BERT-GCN model was evaluated on small datasets of only thousands of papers, partly due to the high cost of computing the GCN, which limited its scalability for recommending citations from large paper databases. Although recent studies \cite{medic-snajder-2020-improved,10.1109/TASLP.2019.2949925,10.1145/3077136.3080730,10.1145/2600428.2609585} adopted the prefetching-reranking strategy to improve the scalability, the prefetch part (BM25 or TF-IDF) only served for creating datasets for training and evaluating the reranking model, since the target cited paper was added manually if it was not retrieved by the prefetch model, i.e. the recall of the target among the candidate papers was set to $1$. Therefore, these recommendation systems were evaluated in an artificial situation with an ideal prefetching model that in reality does not exist.

Supervised methods for citation recommendation rely on the availability of numerous labeled data for training. It is challenging to assemble a dataset for local citation recommendation due to the need of parsing the full text of papers to extract the local contexts and finding citations that are also available in the dataset, which eliminates a large bulk of data. Therefore, existing datasets on local citation recommendation are usually limited in size. For example, the ACL-200 \cite{medic-snajder-2020-improved} and the FullTextPeerRead \cite{jeong2019context} contain only thousands of papers. One of the largest datasets is RefSeer used in Medi{\'c} and {\v{Snajder}} \cite{medic-snajder-2020-improved}, which contains $0.6$ million papers in total, but this dataset is not up-to-date as it only contains papers prior to 2015. 
Although unarXive \cite{saier_unarxive_2020}, a large dataset for citation recommendation, exists, this dataset does not meet the needs of our task because: 1) papers in unarXive are not parsed in a structured manner. For example, the abstract is not separated from the full text, which makes it difficult to construct a global context in our experiments; 2) the citation context is usually a single sentence containing a citation marker, even if the sentence does not contain sufficient contextual information, e.g., ``For details, see [\#]''. These caveats motivate the creation of a novel dataset of high quality. 

\section{Proposed Dataset}\label{sec:new_datasets}
We create a new dataset for local citation recommendation using arXiv papers contained in S2ORC~\cite{lo2020s2orc}, a large-scale scientific paper corpus.
Each paper in S2ORC has an identifier of the paper source, such as arXiv or PubMed. Using this identifier, we first obtain all arXiv papers with available titles and abstracts.  The title and abstract of each paper are required because they are used as the global context of a query from that paper or as a representation of the paper's content when the latter is a candidate to be ranked. From the papers we then extract the local contexts by parsing those papers for which the full text is available: For each reference in the full text, if the cited paper is also available in the arXiv paper database, we replace the reference marker such as ``[\#]'' or ``XXX et al.'' with a special token such as ``CIT'', and collect 200 characters surrounding the replaced citation marker as the local context. Note that we ``cut off'' a word if it lied on the 200-character boundary, following the setting of the ACL-200 and the RefSeer datasets proposed in Medi{\'c} and {\v{Snajder}} \cite{medic-snajder-2020-improved}.

\begin{table}[ht]
\centering
\resizebox{.9\linewidth}{!}{ 
\begin{tabular}{lcccccccccc}
\toprule
\multirow{2}*{Dataset} &$\ \ $& \multicolumn{5}{c}{Number of local contexts} & $\ \ $ & \multirow{2}*{\shortstack{ Number of\\papers } } &$\ \ $&  \multirow{2}*{\shortstack{ publication years } }\\
\cline{3-7}  \\[-1em]
 &&  Train & & Val && Test && && \\\midrule
ACL-200 && $30,390$ && $9,381$ && $9,585$ &&$19,776$ && $2009$ -- $2015$ \\
FullTextPeerRead && $9,363$ && $492$ && $6,814$ &&$4,837$ && $2007$ -- $2017$ \\
RefSeer && $3,521,582$ && $124,911$ && $126,593$ &&$624,957$ && $\ \ \ \ \ \ $ -- $2014$ \\
arXiv (Ours) &&  $2,988,030$ && $112,779$ && $104,401$ &&$1,661,201$ && $1991$ -- $2020$ \\

\bottomrule
\end{tabular}
}
\caption{ \label{tab:datasets_statistics} Statistics of the datasets for local citation recommendation.}  
\end{table}
Table \ref{tab:datasets_statistics} shows the statistics of the created arXiv dataset and the comparison with existing datasets used in this paper. As the most recent contexts available in the arXiv dataset is from April 2020, we use the contexts from 1991 to 2019 as the training set, the contexts from January 2020 to February 2020 as the validating set, and the contexts from March 2020 to April 2020 as the test set. The sizes of the arXiv training, validating, and testing sets are comparable to RefSeer, one of the largest existing datasets, whereas our arXiv dataset contains a much larger number of papers, and there are more recently published papers available in the arXiv dataset. These features make the arXiv dataset a more challenging and up-to-date test bench.

\section{Approach}\label{sec:approach}

Our two-stage telescope citation recommendation system is similar to that of Bhagavatula et al. \cite{bhagavatula-etal-2018-content}, 
composed of a fast \textbf{prefetching} model and a slower \textbf{reranking} model. 

\subsection{Prefetching Model}\label{sec:prefetch}

The prefetching model scores and ranks all papers in the database to fetch a rough initial subset of candidates. We designed a representation-focused ranking model \cite{guo2019deep} that computes a query embedding for each input query and ranks each candidate document according to the cosine similarity between the query embedding and the pre-computed document embedding. 

\begin{figure}
    \centering
    \begin{subfigure}{0.35\textwidth}
      \centering
      \includegraphics[width=.9\linewidth ]{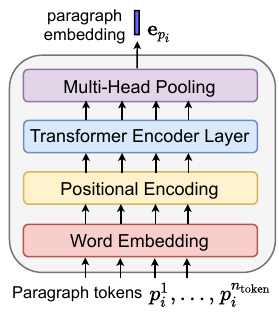}
      \caption{Paragraph Encoder.}
      \label{fig:single_paragraph_encoder}
    \end{subfigure}
    \begin{subfigure}{0.59\textwidth}
      \centering
      \includegraphics[width=.9\linewidth]{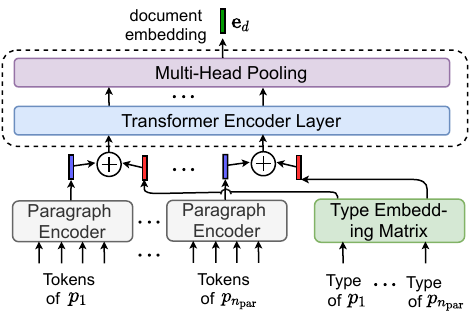}
      \caption{Document Encoder.}
      \label{fig:multiple_paragraph_encoder}
    \end{subfigure}
    \caption{The Hierarchical-Attention text encoder (HAtten) used in the prefetching step is composed of a paragraph encoder (a) and a document encoder (b).}
    \label{fig:prefetch_model}
\end{figure}

The core of the prefetching model is a light-weight text encoder that efficiently computes the embeddings of queries and candidate documents. As shown in Figure \ref{fig:prefetch_model}, the encoder processes each document or query in a two-level hierarchy, consisting of two components: a paragraph encoder and a document encoder. 

\subsubsection{Paragraph Encoder}
\label{para_enc}
For each paragraph $p_i$ in the document, the paragraph encoder (Figure \ref{fig:single_paragraph_encoder}) takes as input the token sequence $p_i = [w_1,\dots,w_{n_\text{i}}]$ composed of $n_\text{i}$ tokens (words)  to output the paragraph embedding $\bm{e}_{p_{i}}$ as a single vector. 
In order to incorporate positional information of the tokens, the paragraph encoder makes use of positional encoding. Contextual information is encoded with a single transformer encoder layer following the configuration in Vaswani et al. \cite{vaswani2017attention}, Figure \ref{fig:single_paragraph_encoder}. To obtain a single fixed-size embedding $\bm{e}_p$ from a variably sized paragraph, the paragraph encoder processes the output of the transformer encoder layer with a multi-head pooling layer \cite{liu2019hierarchical} with trainable weights. Let $x_k \in \mathbb{R}^{d}$ be the output of the transformer encoder layer for token $w_k$ in a paragraph $p_i$. For each head $j \in \{1,\dots, n_\text{head}\}$ in the multi-head pooling layer, we first compute a value vector $v_k^j \in \mathbb{R}^{d/n_\text{head}}$ as well as an attention score $\hat{a}^j_k \in \mathbb{R}$ associated with that value vector:
\begin{align}
v_k^j  &= \text{Linear}^j_v( x_k ),\ \ 
a_k^j = \text{Linear}^j_a( x_k ), \ \ 
\hat{a}_k^j = \frac{ \exp{ a_k^j } }{\sum_{m=1}^{n_\text{token}}{ \exp{ a_m^j } } },
\end{align}
where $\text{Linear}()$ denotes a trainable linear transformation. 
The weighted value vector $\hat{v}^j$ then results from the sum across all value vectors weighed by their corresponding attention scores: $\hat{v}^j = \sum_{m=1}^{n_\text{par}}{\hat{a}_m^j v_m^j }$.
The final paragraph embedding $\bm{e}_p$ is constructed from the weighted value vectors $\hat{v}^j$ of all heads by a ReLU activation \cite{nair2010rectified} followed by a linear transformation:
\begin{align}
    \bm{e}_p &= \text{Linear}_p( \text{ReLU}(\text{Concat}(  \hat{v}^1, \dots, \hat{v}^{n_\text{head}}  ) )).
\end{align}

\subsubsection{Document Encoder}
In order to encode documents with two fields given by the title and the abstract, or to encode queries given by three fields: the local context, the title, and the abstract of the citing paper, we treat each field (local context, title, and abstract) as a paragraph.
For a document of $n_\text{par}$ paragraphs $d =[p_1, \dots, p_{n_{\text{par}}}]$, we first compute the embeddings of all paragraphs $p_i$.

Not all fields and paragraphs are treated equally in our document encoder. To allow the document encoder to distinguish between fields, we introduce a \textit{paragraph type} variable, which refers to the field type from which the paragraph originates. We distinguish between three paragraph types: the title, the abstract, and the local context. Each type is associated with a learnable type embedding that has the same dimension as the paragraph embedding. Inspired by the BERT model \cite{devlin-etal-2019-bert}, we produce a type-aware paragraph embedding by adding the type embedding of the given paragraph to the corresponding paragraph embedding (Figure \ref{fig:multiple_paragraph_encoder}). All type-aware paragraph embeddings are then fed into a transformer encoder layer followed by a multi-head pooling layer (of identical structures as the ones in the paragraph encoder), which then results in the final document embedding~$\bm{e}_{d}$.

\subsubsection{Prefetched document candidates}
\label{prefetch_list}
The prefetched document candidates are found by identifying the $K$ nearest document embeddings to the query embedding in terms of cosine similarity. The ranking is performed using a brute-force nearest neighbor search among all document embeddings as shown in Figure \ref{fig:pipeline}.


\subsection{Reranking Model}

\begin{figure}
    \centering
      \includegraphics[width=0.95\linewidth]{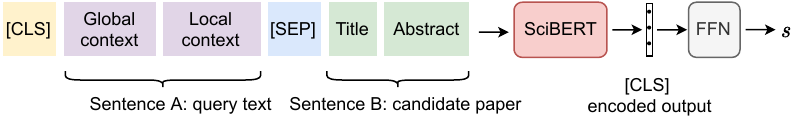}
      \caption{Structure of our SciBERT Reranker.}
      \label{fig:scibert_reranker}
    
\end{figure}

The reranking model performs a fine-grained comparison between a query $q$ (consisting of a local and a global context) and each prefetched document candidate (its title and the abstract). The relevance scores of the candidates constitute the final output of our model. 
We design a reranker based on SciBERT \cite{beltagy2019scibert}, which is a BERT model \cite{devlin-etal-2019-bert} trained on a large-scale corpus of scientific articles. The input of the SciBERT reranker has the following format: ``[CLS] Sentence A [SEP] Sentence B'', where sentence A is the concatenation of the global context (title and abstract of the citing paper) and the local context of the query, and sentence B is the concatenation of the title and the abstract of the candidate paper to be scored,  Figure \ref{fig:scibert_reranker}. The SciBERT-encoded vector for the ``[CLS]'' token is then fed into a feed-forward network that outputs the relevance score $s\in[0,1]$ provided via a sigmoid function. 

\subsection{ Loss Function }
\label{sec:loss_func}
We use a triplet loss both to train our HAtten text encoder for prefetching and to finetune the SciBERT reranker\textbf{}. The triplet loss is based on the similarity $s(q,d)$ between the query $q$ and a document $d$. For the prefetching step, $s(q,d)$ is given by the cosine similarity between the query embedding $v_q$ and the document embedding $v_d$, both computed with the HAtten encoder. For the reranking step, 
$s(q,d)$ is given by the relevance score computed by the SciBERT reranker. 
In order to maximize the relevance score between the query $q$ and the cited document $d_+$ (the positive pair $(q,d_+)$) and to minimize the score between $q$ and any non-cited document $d_\_$ (a negative pair $(q, d_\_)$), we minimize the triplet loss:
\begin{align}
    \mathcal{L} = \text{max}\large [ s( q, d_{\_} ) - s(q, d_{+}) + m, 0   \large]
    \label{eq:triplet_loss}
\end{align}
where the margin $m>0$ sets the span over which the loss is sensitive to the similarity of negative pairs.

For fast convergence during training, it is important to select effective triplets for which $\mathcal{L}$ in Equation \eqref{eq:triplet_loss} is non-zero \cite{7298682}, which is particularly relevant for the prefetching model, since for each query there is only a single positive document but millions of negative documents (e.g., on the arXiv dataset). Therefore, we employ negative and positive mining strategies to train our HAtten encoder, described as follows.

\noindent\textbf{Negative mining} Given a query $q$, we use HAtten's current checkpoint to prefetch the top $K_n$ candidates excluding the cited paper. The HAtten embedding of these prefetched non-cited candidates have high cosine similarity to the HAtten embedding of the query. To increase the similarity between the query and the cited paper while suppressing the similarity between the query and these non-cited candidates, we use the cited paper as the positive document and select the negative document from these $K_n$ overly similar candidates.

\noindent\textbf{Positive mining}
Among the prefetched non-cited candidates, the documents with objectively high textual similarity (e.g. measured by word overlapping, such as the Jaccard index \cite{bhagavatula-etal-2018-content}) to the query were considered relevant to the query, even if they were not cited. These textually relevant candidate documents should have a higher cosine similarity to the query than randomly selected documents. Therefore, in parallel with the negative mining strategy, we also select positive documents from the set of textually relevant candidates and select negative documents by random sampling from the entire dataset.

The checkpoint of the HAtten model is updated every $N_\text{iter}$ training iterations, at which point the prefetched non-cited and the textually relevant candidates for negative and positive mining are updated as well. 

In contrast, when fine-tuning SciBERT for reranking, the reranker only needs to rerank the top $K_r$ prefetched candidates. This allows for a simpler triplet mining strategy, which is to select the cited paper as the positive document and randomly selecting a prefetched non-cited papers as the negative document.


\section{Experiments} 
\noindent\textbf{Implementation Details}
In the prefetching step, we used as word embeddings of the HAtten text encoder the pre-trained 200-dimensional GloVe embeddings \cite{pennington-etal-2014-glove}, which were kept fixed during training.
There are $64$ queries in a mini-batch, each of which was accompanied by $1$ cited paper, $4$ non-cited papers randomly sampled from the top $K_n=100$ prefetched candidates, and $1$ randomly sampled paper from the whole database, which allow us to do negative and positive mining with the
mini-batch as described in Section \ref{sec:loss_func}. The HAtten's checkpoint was updated every $N_\text{iter}=5000$ training iterations.

In the reranking step, we initialized the SciBERT reranker with the pretrained model provided in Beltagy et al. \cite{beltagy2019scibert}. The feed-forward network in Figure \ref{fig:scibert_reranker} consisting of a single linear layer was randomly initialized. Within a mini-batch there was $1$ query, $1$ cited paper (positive sample), and $62$ documents (negative samples) randomly sampled from the top $K_r=2000$ prefetched non-cited documents. In the triplet loss function the margin $m$ was set to~$0.1$.

We used the Adam optimizer \cite{kingma2014adam} with $\beta_1=0.9$ and $\beta_2=0.999$. In the prefetching step, the learning rate was set to $\alpha=1e^{-4}$ and the weight decay to $1e^{-5}$, while in the reranking step these were set to $1e^{-5}$ and to $1e^{-2}$ for fine-tuning SciBERT, respectively. The models were trained on eight NVIDIA GeForce RTX 2080 Ti 11GB GPUs and tested on two Quadro RTX 8000 GPUs. 

\noindent\textbf{Evaluation Metrics}
We evaluated the recommendation performance using the Mean Reciprocal Rank (MRR) \cite{Voorhees99thetrec-8} and the Recall@K (R@K for short), consistent with previous work \cite{medic-snajder-2020-improved,10.1145/3383583.3398534,jeong2019context}. The MRR measures the reciprocal rank of the actually cited paper among the recommended candidates, averaged over multiple queries. The R@K evaluates the percentage of the cited paper appearing in the top $K$ recommendations.

\noindent\textbf{Baselines} 
In the prefetching step, we compare our HAtten with the following baselines: BM25, Sent2vec \cite{pagliardini2017unsupervised}, and NNSelect \cite{bhagavatula-etal-2018-content}. BM25 was used as the prefetching method in previous works \cite{medic-snajder-2020-improved,10.1145/3077136.3080730,10.1145/2600428.2609585}. Sent2vec is an unsupervised text encoder which computes a text embedding by averaging the embeddings of all words in the text. We use the 600-dim Sent2vec pretrained on Wikipedia. NNSelect \cite{bhagavatula-etal-2018-content} computes text embeddings also by averaging, and the trainable parameters are the magnitudes of word embeddings that we trained on each dataset using the same training configuration as our HAtten model. 

In the reranking step, we compare our fine-tuned SciBERT reranker with the following baselines: 1) a Neural Citation Network (NCN) with an encoder-decoder architecture \cite{10.1145/3077136.3080730,Frber2020NeuralCR}; 2) DualEnh and DualCon \cite{medic-snajder-2020-improved} that score each candidate using both semantic information and bibliographic information and 3) BERT-GCN \cite{jeong2019context}.
Furthermore, to analyze the influence on ranking performance of diverse pretraining corpuses for BERT, we compared our SciBERT reranker with a BERT reranker that was pretrained on a non-science specific corpus \cite{devlin-etal-2019-bert} and then fine-tuned on the reranking task. 

For a fair performance comparison of our reranker with those of other works, we adopted the prefetching strategies from each of these works. On ACL-200 and RefSeer, we tested our SciBERT reranker on the test sets provided in Medi{\'c} and {\v{Snajder}} \cite{medic-snajder-2020-improved}. For each query in the test set, we prefetched $n$ ($n=2000$ for ACL-200 and $n=2048$ for RefSeer) candidates using BM25, and manually added the cited paper as candidate if it was not found by BM25. In other words, we constructed our test set using an ``oracle-BM25'' with $\text{R}@n=1$. On the FullTextPeerRead dataset, we used our SciBERT reranker to rank all papers in the database without prefetching, in line with the setting in BERT-GCN~\cite{jeong2019context}. On our newly proposed arXiv dataset, we fetched the top 2000 candidates for each query in the test set using the ``oracle-BM25'' as introduced above. 

\section{Results and Discussion}
In this section, we first present the evaluation results of our prefetching and reranking models separately and compare them with baselines. Then, we evaluate the performance of the entire prefetching-reranking pipeline, and analyze the influence of the number of prefetched candidates to be reranked on the overall recommendation performance.

\subsection{Prefetching Results}
\begin{table}
\centering
\resizebox{\linewidth}{!}{ 
\begin{tabular}{lllcccccccccccccccc}
\toprule
 \multirow{1}*{Dataset} && \multirow{1}*{Model} && avg. prefetch time (ms) && MRR  &&  R@10  &&  R@100 &&  R@200 &&  R@500 &&  R@1000 &&  R@2000\\ \midrule
\multirow{4}*{ACL-200} && BM25 && $9.9\pm 20.1$  && 0.138&&0.263&&0.520&&0.604&&0.712&&0.791&&0.859 \\
&& Sent2vec&& $1.8\pm19.5$&& 0.066&&0.127&&0.323&&0.407&&0.533&&0.640&&0.742\\
&& NNSelect&&$1.8\pm3.8$ && 0.076&&0.150&&0.402&&0.498&&0.631&&0.722&&0.797\\
&& HAtten && $2.7\pm3.8$ &&\textbf{0.148}*&&\textbf{0.281}*&&\colorbox{red}{\textbf{0.603}*}&&\colorbox{red}{\textbf{0.700}*}&&\colorbox{red}{\textbf{0.803}*}&&\colorbox{red}{\textbf{0.870}*}&&\colorbox{red}{\textbf{0.924}*}\\\midrule
\multirow{4}*{\shortstack{FullText-\\PeerRead}}&&BM25&& $5.1\pm18.6$ &&\textbf{0.185}*&&\textbf{0.328}*&&0.609&&0.694&&0.802&&0.877&&0.950\\
&&Sent2vec&& $1.7\pm19.6$ &&0.121&&0.215&&0.462&&0.561&&0.694&&0.794&&0.898\\
&&NNSelect&& $1.7\pm4.8$ &&0.130&&0.255&&0.572&&0.672&&0.790&&0.869&&0.941\\
&&HAtten&& $2.6\pm4.9$ &&0.167&&0.306&&\textbf{0.649}*&&\colorbox{red}{\textbf{0.750}*}&&\colorbox{red}{\textbf{0.870}*}&&\colorbox{red}{\textbf{0.931}*}&&\colorbox{red}{\textbf{0.976}*}\\\midrule
\multirow{4}*{RefSeer} &&BM25&&$216.2\pm84.9$ && 0.099&&0.189&&0.398&&0.468&&0.561&&0.631&&0.697\\
&&Sent2vec&& $6.0\pm20.9$ &&0.061&&0.111&&0.249&&0.306&&0.389&&0.458&&0.529\\
&&NNSelect&& $4.3\pm5.5$ &&0.044&&0.080&&0.197&&0.250&&0.331&&0.403&&0.483\\
&&HAtten&& $6.2\pm7.3$ &&\textbf{0.115}*&&\textbf{0.214}*&&\colorbox{red}{\textbf{0.492}*}&&\colorbox{red}{\textbf{0.589}*}&&\colorbox{red}{\textbf{0.714}*}&&\colorbox{red}{\textbf{0.795}*}&&\colorbox{red}{\textbf{0.864}*}\\\midrule
\multirow{4}*{arXiv}&&BM25&& $702.2\pm104.7$ &&0.118&&0.222&&0.451&&0.529&&0.629&&0.700&&0.763\\
&&Sent2vec&& $11.3\pm13.6$ &&0.072&&0.131&&0.287&&0.347&&0.435&&0.501&&0.571\\
&&NNSelect&& $6.9\pm4.6$ &&0.042&&0.079&&0.207&&0.266&&0.359&&0.437&&0.520\\
&&HAtten&& $8.0\pm4.5$ &&\textbf{0.124}*&&\textbf{0.241}*&&\colorbox{red}{\textbf{0.527}*}&&\colorbox{red}{\textbf{0.619}*}&&\colorbox{red}{\textbf{0.734}*}&&\colorbox{red}{\textbf{0.809}*}&&\colorbox{red}{\textbf{0.871}*}\\
\bottomrule
\end{tabular}
}
\caption{ \label{tab:res_prefetch} Prefetching performance. For Tables \ref{tab:res_prefetch}-\ref{tab:whole_pipeline}, the asterisks ``*'' indicate statistical significance ($p<0.05$) in comparison with the closest baseline in a t-test. The \colorbox{red}{red} color indicates a 
large ($>0.8$) Cohen's d effect size \cite{cohen2013statistical}.
}
\end{table}

Our HAtten model significantly outperformed all baselines (including the strong baseline BM25, Table \ref{tab:res_prefetch}) on the ACL-200, RefSeer and the arXiv datasets, evaluated in terms of MRR and R@K. We observed that, first, for larger $K$, such as $K=200,500,1000,2000$, the improvement of R@K with respect to the baselines is more pronounced on all four datasets, where the increase is usually larger than $0.1$, which means that the theoretical upper bound of the final reranking recall will be higher when using our HAtten prefetching system. Second, the improvements of R@K on large datasets such as RefSeer and arXiv are more prominent than on small datasets such as ACL-200 and FullTextPeerRead, which fits well with the stronger need of a prefetching-reranking pipeline on large datasets due to the speed-accuracy tradeoff.

The advantage of our HAtten model is also reflected in the average prefetching time. As shown in Table \ref{tab:res_prefetch}, the HAtten model shows faster prefetching than BM25 on large datasets such as RefSeer and arXiv. This is because for HAtten, both text encoding and embedding-based nearest neighbor search can be accelerated by GPU computing, while BM25\footnote{We implemented the Okapi BM25 \cite{manning2008introduction}, with $k=1.2, b=0.75$.} benefits little from GPU acceleration because it is not vector-based. Although other embedding-based baselines such as Sent2vec and NNSelect also exhibit fast prefetching, our HAtten prefetcher has advantages in terms of both speed and accuracy.
\subsection{Reranking Results}
\label{sec:reranking_res}

\begin{table}
\centering
\resizebox{\linewidth}{!}{ 
\begin{tabular}{llccccccccccccccccccccccc}
\toprule
 \multirow{2}*{Model} && && \multicolumn{3}{c}{ ACL-200} && &&  \multicolumn{3}{c}{\shortstack{FullTextPeerRead}} && && \multicolumn{3}{c}{RefSeer} && && \multicolumn{3}{c}{arXiv}\\ 
 \cline{5-7}  \cline{11-13}  \cline{17-19}  \cline{23-25} \\[-1em]
  && && MRR && R@10 && && MRR && R@10 && && MRR && R@10 && && MRR && R@10 
 \\\midrule
 
 NCN    && && - && - && && - && - && && 0.267 && 0.291 && && - && - \\
 DualCon   && && 0.335 && 0.647 && && - && - && && 0.206 && 0.406 && && - && - \\
 DualEnh  && && 0.366 && 0.703 && && - && - && && 0.280 && 0.534 && && - && - \\
 BERT-GCN  && && - && - && && 0.418 && 0.529 && && - && - && && - && - \\
 BERT Reranker    && && 0.482 && 0.736 && && 0.458 && 0.706 && && 0.309 && 0.535 && &&  0.226 && 0.399 \\
 SciBERT Reranker   && && \colorbox{red}{\textbf{0.531}*} &&\colorbox{red}{ \textbf{0.779}*} && && \colorbox{red}{\textbf{0.536}*} && \colorbox{red}{\textbf{0.773}*} && &&  \colorbox{red}{\textbf{0.380}*} &&  \colorbox{red}{\textbf{0.623}*}  && &&  \colorbox{red}{\textbf{0.278}*}  && \colorbox{red}{\textbf{0.475}*}  \\
\bottomrule
\end{tabular}
}
\caption{ \label{tab:res_reranking} Comparison of reranking performance on four datasets.}  
\end{table}

As shown in Table \ref{tab:res_reranking}, the SciBERT reranker significantly outperformed previous state-of-the-art models on the ACL-200, the RefSeer, and the FullTextPeerRead datasets. We ascribe this improvement to BERT's ability of capturing the semantic relevance between the query text and the candidate text, which is inherited from the ``next sentence prediction'' pretraining task that aims to predict if two sentences are consecutive. The SciBERT reranker also performed significantly better than its BERT counterpart, suggesting that large language models pretrained on scientific papers' corpus are advantageous for citation reranking.
\subsection{Performance of entire Recommendation Pipeline}

\begin{table}
\centering
\resizebox{.85\linewidth}{!}{ 
\begin{tabular}{llllllccccccccc}
\toprule
 \multirow{2}*{Dataset} && \multicolumn{3}{c}{Recommendation Pipeline} && \multicolumn{9}{c}{Number of reranked candidates} \\
 \cline{3-5} \cline{7-15} \\[-1em]
  && \multicolumn{1}{c}{$\ \ \ $Prefetch}$\ \ \ $ && \multicolumn{1}{c}{Rerank} && $\ \ \ $100$\ \ \ $ &&  $\ \ \ $200$\ \ \ $   &&  $\ \ \ $500$\ \ \ $  &&  $\ \ \ $1000$\ \ \ $  &&  $\ \ \ $2000$\ \ \ $  
 \\ \midrule
 
\multirow{2}*{ACL-200} && $\ \ \ $BM25 && SciBERT$_\text{BM25}$  && 0.457    &&   0.501    &&   0.549 &&   0.577    &&    0.595    \\
&& $\ \ \ $HAtten && SciBERT$_\text{HAtten}$  &&  \textbf{0.513}*   &&  \textbf{0.560}*    &&   \textbf{0.599}*    &&  \textbf{ 0.619}*    &&    \textbf{0.633}*  \\\midrule

\multirow{2}*{\shortstack{FullText-\\PeerRead}} && $\ \ \ $BM25 && SciBERT$_\text{BM25}$  && 0.527    &&   0.578    &&   0.639    &&    0.680    &&    0.720   \\
&& $\ \ \ $HAtten && SciBERT$_\text{HAtten}$  &&  \textbf{0.586}*    &&  \colorbox{red}{ \textbf{0.651}*}   &&   \colorbox{red}{\textbf{0.713}*}    &&   \colorbox{red}{\textbf{0.739}*}   &&    \textbf{0.757}*  \\\midrule

\multirow{2}*{\shortstack{RefSeer}} && $\ \ \ $BM25 && SciBERT$_\text{BM25}$  && 0.305    &&   0.332    &&   0.365    &&    0.380    &&    0.383     \\
&& $\ \ \ $HAtten && SciBERT$_\text{HAtten}$  &&   \colorbox{red}{\textbf{0.362}*}   &&   \colorbox{red}{\textbf{0.397}*}    &&   \colorbox{red}{\textbf{0.428}*}    &&   \colorbox{red}{\textbf{0.443}*}    &&   \colorbox{red}{\textbf{0.454}*}    \\\midrule

\multirow{2}*{\shortstack{arXiv}} && $\ \ \ $BM25 && SciBERT$_\text{BM25}$  &&  0.333    &&   0.357    &&   0.377    &&    0.389    &&    0.391    \\
&& $\ \ \ $HAtten && SciBERT$_\text{HAtten}$  &&    \textbf{0.374}*    &&   \colorbox{red}{\textbf{0.397}*}    &&  \colorbox{red}{\textbf{0.425}*}    &&    \colorbox{red}{\textbf{0.435}*}    &&    \colorbox{red}{\textbf{0.439}*}      \\

\bottomrule
\end{tabular}
}
\caption{ \label{tab:whole_pipeline} The performance of the entire prefetching-reranking pipeline, measured in terms of R@10 of the final reranked document list. We varied the number of prefetched candidates for reranking. For the RefSeer and arXiv datasets, we evaluated performance on a subset of 10K examples from the test set due to computational resource limitations. }
\end{table}

The evaluation in Section \ref{sec:reranking_res} only reflects the reranking performance because the prefetched candidates are obtained by an oracle-BM25 that guarantees inclusion of the cited paper among the prefetched candidates, even though such an oracle prefetching model does not exist in reality. Evaluating recommendation systems in this context risks overestimating the performance of the reranking part and underestimating the importance of the prefetching step. 
To better understand the recommendation performance in real-world scenarios, we compared two pipelines: 1) BM25 prefetching + SciBERT reranker  fine-tuned on BM25-prefetched candidates, denoted as SciBERT$_\text{BM25}$; 2) HAtten prefetching + SciBERT$_\text{HAtten}$ reranker fine-tuned on HAtten-prefetched candidates. 
We evaluated recommendation performance by R@10 of the final reranked document list and monitored the dependence of R@10 on the number of prefetched candidates for reranking. 

\begin{figure}[ht]
    \centering

    \includegraphics[width=.75\linewidth]{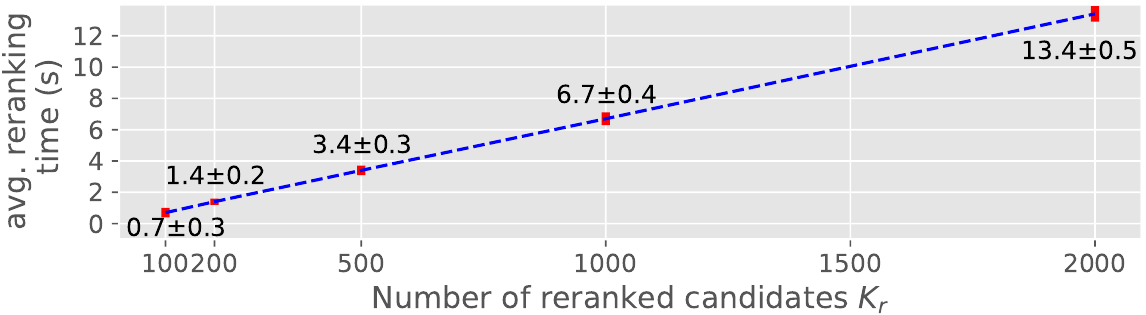}
      
    \caption{The reranking time of the SciBERT reranker linearly increases with the number of reranked candidates $K_r$, tested on arXiv. In comparison, the prefetching time is invariant of $K_r$, as the prefetcher always scores and ranks all documents in the database to fetch the candidates to be reranked.}
    \label{fig:rerank_time}
\end{figure}

As shown in Table \ref{tab:whole_pipeline}, the HAtten-based pipeline achieves competitive performance, even when compared with the oracle prefetching model in Section \ref{sec:reranking_res}.
In particular, on the FullTextPeerRead dataset, using our HAtten-based pipeline, we only need to rerank 100 prefetched candidates to outperform the BERT-GCN model (Table \ref{tab:res_reranking}) that reranked all 4.8k papers in the database.


Compared to the BM25-based pipeline, our HAtten-based pipeline achieves significantly higher R@10 for any given number of prefetched candidates. Our reranker needs to rerank only 200 to 500 candidates to match the recall score of the BM25-based pipeline needing to rerank 2000 candidates. For large datasets like RefSeer and arXiv, such improvements are even more pronounced.
Our pipeline achieves a much higher throughput.  For example, on the arXiv dataset, in order to achieve an overall R@10=$0.39$, the BM25-based pipeline takes 0.7 s (Table \ref{tab:res_prefetch}) to prefetch 2000 candidates and it takes another 13.4 s (Figure \ref{fig:rerank_time}) to rerank them, which in total amounts to 14.1 s. In contrast, the HAtten-based pipeline only takes 8 ms to prefetch 200 candidates and 1.4 s to rerank them, which amounts to 1.4 s. This results in a 90\% reduction of overall recommendation time achieved by our pipeline.

These findings provide clear evidence that a better-performing prefetching model is critical to a large-scale citation recommendation pipeline, as it allows the reranking model to rerank fewer candidates while maintaining recommendation performance, resulting in a better speed-accuracy tradeoff.

\section{Conclusion}
The speed-accuracy tradeoff is crucial for evaluating recommendation systems in real-world settings. While reranking models have attracted increasing attention for their ability to improve recall and MRR scores, in this paper we show that it is equally important to design an efficient and accurate prefetching system. 
In this regard, we propose the  HAtten-SciBERT recommendation pipeline, in which our HAtten model effectively prefetches a list of candidates with significantly higher recall than the baseline, which allows our fine-tuned SciBERT-based reranker to operate on fewer candidates with better speed-accuracy tradeoff.
Furthermore, by releasing our large-scale arXiv-based dataset, we provide a new testbed for research on local citation recommendation in real-world scenarios.

\section*{Acknowledgements}
We acknowledge support from the Swiss National
Science Foundation (grant 31003A\_182638) and the NCCR Evolving Language, Swiss National Science Foundation Agreement No. 51NF40\_180888. We also thank the anonymous reviewers for their useful comments.

\bibliographystyle{splncs04}
\bibliography{mybibliography}

\begin{thebibliography}{10}
\providecommand{\url}[1]{\texttt{#1}}
\providecommand{\urlprefix}{URL }
\providecommand{\doi}[1]{https://doi.org/#1}

\bibitem{beltagy2019scibert}
Beltagy, I., Lo, K., Cohan, A.: {S}ci{BERT}: A pretrained language model for
  scientific text. In: Proceedings of the 2019 Conference on Empirical Methods
  in Natural Language Processing and the 9th International Joint Conference on
  Natural Language Processing (EMNLP-IJCNLP). pp. 3615--3620. Association for
  Computational Linguistics, Hong Kong, China (Nov 2019).
  \doi{10.18653/v1/D19-1371}, \url{https://www.aclweb.org/anthology/D19-1371}

\bibitem{bhagavatula-etal-2018-content}
Bhagavatula, C., Feldman, S., Power, R., Ammar, W.: Content-based citation
  recommendation. In: Proceedings of the 2018 Conference of the North
  {A}merican Chapter of the Association for Computational Linguistics: Human
  Language Technologies, Volume 1 (Long Papers). pp. 238--251. Association for
  Computational Linguistics, New Orleans, Louisiana (Jun 2018).
  \doi{10.18653/v1/N18-1022}, \url{https://www.aclweb.org/anthology/N18-1022}

\bibitem{cohen2013statistical}
Cohen, J.: Statistical power analysis for the behavioral sciences. Academic
  press (2013)

\bibitem{10.1109/TASLP.2019.2949925}
Dai, T., Zhu, L., Wang, Y., Carley, K.M.: Attentive stacked denoising
  autoencoder with bi-lstm for personalized context-aware citation
  recommendation. IEEE/ACM Trans. Audio, Speech and Lang. Proc.  \textbf{28},
  553–568 (Jan 2020). \doi{10.1109/TASLP.2019.2949925},
  \url{https://doi.org/10.1109/TASLP.2019.2949925}

\bibitem{devlin-etal-2019-bert}
Devlin, J., Chang, M.W., Lee, K., Toutanova, K.: {BERT}: Pre-training of deep
  bidirectional transformers for language understanding. In: Proceedings of the
  2019 Conference of the North {A}merican Chapter of the Association for
  Computational Linguistics: Human Language Technologies, Volume 1 (Long and
  Short Papers). pp. 4171--4186. Association for Computational Linguistics,
  Minneapolis, Minnesota (Jun 2019). \doi{10.18653/v1/N19-1423},
  \url{https://www.aclweb.org/anthology/N19-1423}

\bibitem{10.1145/3077136.3080730}
Ebesu, T., Fang, Y.: Neural citation network for context-aware citation
  recommendation. In: Proceedings of the 40th International ACM SIGIR
  Conference on Research and Development in Information Retrieval. p.
  1093–1096. SIGIR '17, Association for Computing Machinery, New York, NY,
  USA (2017). \doi{10.1145/3077136.3080730},
  \url{https://doi.org/10.1145/3077136.3080730}

\bibitem{Frber2020NeuralCR}
F{\"a}rber, M., Klein, T., Sigloch, J.: Neural citation recommendation: A
  reproducibility study. In: BIR@ECIR (2020)

\bibitem{10.1145/3383583.3398534}
F\"{a}rber, M., Sampath, A.: Hybridcite: A hybrid model for context-aware
  citation recommendation. In: Proceedings of the ACM/IEEE Joint Conference on
  Digital Libraries in 2020. p. 117–126. JCDL '20, Association for Computing
  Machinery, New York, NY, USA (2020). \doi{10.1145/3383583.3398534},
  \url{https://doi.org/10.1145/3383583.3398534}

\bibitem{F_rber_2020}
Färber, M., Jatowt, A.: Citation recommendation: approaches and datasets.
  International Journal on Digital Libraries  \textbf{21}(4),  375–405 (Aug
  2020). \doi{10.1007/s00799-020-00288-2},
  \url{http://dx.doi.org/10.1007/s00799-020-00288-2}

\bibitem{gokcce2020embedding}
G{\"o}k{\c{c}}e, O., Prada, J., Nikolov, N.I., Gu, N., Hahnloser, R.H.:
  Embedding-based scientific literature discovery in a text editor application.
  In: Proceedings of the 58th Annual Meeting of the Association for
  Computational Linguistics: System Demonstrations. pp. 320--326. Association
  for Computational Linguistics, Online (Jul 2020).
  \doi{10.18653/v1/2020.acl-demos.36},
  \url{https://www.aclweb.org/anthology/2020.acl-demos.36}

\bibitem{guo2019deep}
Guo, J., Fan, Y., Pang, L., Yang, L., Ai, Q., Zamani, H., Wu, C., Croft, W.B.,
  Cheng, X.: A deep look into neural ranking models for information retrieval.
  Information Processing \& Management p. 102067 (2019)

\bibitem{he2010context}
He, Q., Pei, J., Kifer, D., Mitra, P., Giles, L.: Context-aware citation
  recommendation. In: Proceedings of the 19th international conference on World
  wide web. pp. 421--430 (2010)

\bibitem{herdan1960type}
Herdan, G.: Type-token mathematics, vol.~4. Mouton (1960)

\bibitem{huang2012recommending}
Huang, W., Kataria, S., Caragea, C., Mitra, P., Giles, C.L., Rokach, L.:
  Recommending citations: translating papers into references. In: Proceedings
  of the 21st ACM international conference on Information and knowledge
  management. pp. 1910--1914 (2012)

\bibitem{hunter2006biomedical}
Hunter, L., Cohen, K.B.: Biomedical language processing: what's beyond pubmed?
  Molecular cell  \textbf{21}(5),  589--594 (2006)

\bibitem{jeong2019context}
Jeong, C., Jang, S., Park, E.L., Choi, S.: A context-aware citation
  recommendation model with {BERT} and graph convolutional networks.
  Scientometrics  \textbf{124}(3),  1907--1922 (2020).
  \doi{10.1007/s11192-020-03561-y},
  \url{https://doi.org/10.1007/s11192-020-03561-y}

\bibitem{kingma2014adam}
Kingma, D.P., Ba, J.: Adam: {A} method for stochastic optimization. In: Bengio,
  Y., LeCun, Y. (eds.) 3rd International Conference on Learning
  Representations, {ICLR} 2015, San Diego, CA, USA, May 7-9, 2015, Conference
  Track Proceedings (2015), \url{http://arxiv.org/abs/1412.6980}

\bibitem{DBLP:conf/iclr/KipfW17}
Kipf, T.N., Welling, M.: Semi-supervised classification with graph
  convolutional networks. In: 5th International Conference on Learning
  Representations, {ICLR} 2017, Toulon, France, April 24-26, 2017, Conference
  Track Proceedings. OpenReview.net (2017),
  \url{https://openreview.net/forum?id=SJU4ayYgl}

\bibitem{10.1145/3197026.3197059}
Kobayashi, Y., Shimbo, M., Matsumoto, Y.: Citation recommendation using
  distributed representation of discourse facets in scientific articles. In:
  Proceedings of the 18th ACM/IEEE on Joint Conference on Digital Libraries. p.
  243–251. JCDL '18, Association for Computing Machinery, New York, NY, USA
  (2018). \doi{10.1145/3197026.3197059},
  \url{https://doi.org/10.1145/3197026.3197059}

\bibitem{liu2019hierarchical}
Liu, Y., Lapata, M.: Hierarchical transformers for multi-document
  summarization. In: Proceedings of the 57th Annual Meeting of the Association
  for Computational Linguistics. pp. 5070--5081. Association for Computational
  Linguistics, Florence, Italy (Jul 2019). \doi{10.18653/v1/P19-1500},
  \url{https://www.aclweb.org/anthology/P19-1500}

\bibitem{10.1145/2600428.2609585}
Livne, A., Gokuladas, V., Teevan, J., Dumais, S.T., Adar, E.: Citesight:
  Supporting contextual citation recommendation using differential search. In:
  Proceedings of the 37th International ACM SIGIR Conference on Research \&
  Development in Information Retrieval. p. 807–816. SIGIR '14, Association
  for Computing Machinery, New York, NY, USA (2014).
  \doi{10.1145/2600428.2609585}, \url{https://doi.org/10.1145/2600428.2609585}

\bibitem{lo2020s2orc}
Lo, K., Wang, L.L., Neumann, M., Kinney, R., Weld, D.S.: S2orc: The semantic
  scholar open research corpus. In: Proceedings of the 58th Annual Meeting of
  the Association for Computational Linguistics. pp. 4969--4983 (2020)

\bibitem{manning2008introduction}
Manning, C.D., Raghavan, P., Schütze, H.: Introduction to Information
  Retrieval. Cambridge University Press, Cambridge, UK (2008),
  \url{http://nlp.stanford.edu/IR-book/information-retrieval-book.html}

\bibitem{medic-snajder-2020-improved}
Medi{\'c}, Z., Snajder, J.: Improved local citation recommendation based on
  context enhanced with global information. In: Proceedings of the First
  Workshop on Scholarly Document Processing. pp. 97--103. Association for
  Computational Linguistics, Online (Nov 2020).
  \doi{10.18653/v1/2020.sdp-1.11}, \url{https://aclanthology.org/2020.sdp-1.11}

\bibitem{nair2010rectified}
Nair, V., Hinton, G.E.: Rectified linear units improve restricted boltzmann
  machines. In: ICML (2010)

\bibitem{nallapati2008joint}
Nallapati, R.M., Ahmed, A., Xing, E.P., Cohen, W.W.: Joint latent topic models
  for text and citations. In: Proceedings of the 14th ACM SIGKDD international
  conference on Knowledge discovery and data mining. pp. 542--550 (2008)

\bibitem{pagliardini2017unsupervised}
Pagliardini, M., Gupta, P., Jaggi, M.: Unsupervised learning of sentence
  embeddings using compositional n-gram features. In: Proceedings of the 2018
  Conference of the North {A}merican Chapter of the Association for
  Computational Linguistics: Human Language Technologies, Volume 1 (Long
  Papers). pp. 528--540. Association for Computational Linguistics, New
  Orleans, Louisiana (Jun 2018). \doi{10.18653/v1/N18-1049},
  \url{https://www.aclweb.org/anthology/N18-1049}

\bibitem{pennington-etal-2014-glove}
Pennington, J., Socher, R., Manning, C.: {G}lo{V}e: Global vectors for word
  representation. In: Proceedings of the 2014 Conference on Empirical Methods
  in Natural Language Processing ({EMNLP}). pp. 1532--1543. Association for
  Computational Linguistics, Doha, Qatar (Oct 2014). \doi{10.3115/v1/D14-1162},
  \url{https://aclanthology.org/D14-1162}

\bibitem{ramos2003using}
Ramos, J., et~al.: Using tf-idf to determine word relevance in document
  queries. In: Proceedings of the first instructional conference on machine
  learning. vol.~242, pp. 133--142. New Jersey, USA (2003)

\bibitem{robertson2009probabilistic}
Robertson, S., Zaragoza, H.: The probabilistic relevance framework: BM25 and
  beyond. Now Publishers Inc (2009)

\bibitem{saier_unarxive_2020}
Saier, T., Färber, M.: {unarXive}: a large scholarly data set with
  publications’ full-text, annotated in-text citations, and links to
  metadata. Scientometrics  \textbf{125}(3),  3085--3108 (Dec 2020).
  \doi{10.1007/s11192-020-03382-z},
  \url{https://doi.org/10.1007/s11192-020-03382-z}

\bibitem{7298682}
Schroff, F., Kalenichenko, D., Philbin, J.: Facenet: A unified embedding for
  face recognition and clustering. In: 2015 IEEE Conference on Computer Vision
  and Pattern Recognition (CVPR). pp. 815--823 (2015).
  \doi{10.1109/CVPR.2015.7298682}

\bibitem{strohman2007recommending}
Strohman, T., Croft, W.B., Jensen, D.: Recommending citations for academic
  papers. In: Proceedings of the 30th annual international ACM SIGIR conference
  on Research and development in information retrieval. pp. 705--706 (2007)

\bibitem{vaswani2017attention}
Vaswani, A., Shazeer, N., Parmar, N., Uszkoreit, J., Jones, L., Gomez, A.N.,
  Kaiser, {\L}., Polosukhin, I.: Attention is all you need. In: Advances in
  neural information processing systems. pp. 5998--6008 (2017)

\bibitem{Voorhees99thetrec-8}
Voorhees, E.M.: The trec-8 question answering track report. In: In Proceedings
  of TREC-8. pp. 77--82 (1999)

\end{thebibliography}

\end{document}